\documentclass[twocolumn,showpacs,prb]{revtex4}

\usepackage{graphicx}%
\usepackage{dcolumn}
\usepackage{amsmath}
\usepackage{amssymb}

\makeatletter
\def\btt#1{\texttt{\@backslashchar#1}}%
\DeclareRobustCommand\bblash{\btt{\@backslashchar}}%
\makeatother

\begin{document}

\title[Short Title]{Josephson Spin Current in Triplet Superconductor Junctions}

\author{Yasuhiro Asano}
\affiliation{
Department of Applied Physics, Hokkaido University, Sapporo 060-8628, Japan}%

\date{\today}

\begin{abstract}
This paper theoretically discusses the spin current in spin-triplet superconductor / 
insulator / spin-triplet superconductor junctions.
At low temperatures, a midgap Andreev resonant state anomalously enhances not only 
the charge current but also the spin current. 
The coupling between the Cooper pairs and the electromagnetic fields leads to the 
Frounhofer pattern in the direct current spin flow 
in magnetic fields and the alternative spin current under applied bias-voltages. 
\end{abstract}

\maketitle
Although the supercurrent usually refers to the dissipationless charge flow in superconductors
or the mass flow in superfluid He, the spin flow carried by spin-triplet Cooper pairs 
is also undoubtedly a supercurrent~\cite{leggett}. 
These supercurrents have a common feature; the spatial gradient of the order parameter 
drives the supercurrent.
It is well known that the charge and mass supercurrents are possible under the spatial 
gradient in the macroscopic phase $\varphi$. 
On the other hand, the spatial gradient of $\boldsymbol{d}$-vector causes the spin 
supercurrent.
In the mean-field theory of superconductivity, $\boldsymbol{d}$ characterizes the
order parameter of spin-triplet pairs as $\hat{\Delta}(\vec{r}\,)=
i\boldsymbol{d}(\vec{r}\,)\cdot\hat{\boldsymbol{\sigma}}\hat{\sigma}_2e^{i\varphi}$
with $\hat{\sigma}_j$ for $j=1-3$ are the Pauli's matrices.
In $^3$He, the angular momentum vector of a Cooper pair points 
the direction normal to the surrounding wall. 
The spin current is expected near the curved wall 
because $\boldsymbol{d}$ and the momentum vector
align to each other due to the dipole-dipole interaction. 
Through the spin dynamics in superfluids, the spin current is detected by 
nuclear magnetic resonance experiments~\cite{bunkov,vuorio,fomin}. 

In bulk superconductors, however, the spin current is usually not expected because
the weak spin anisotropy fixes 
$\boldsymbol{d}$ homogeneously in a certain direction of the crystal lattice.
So far the generation of the spin flow has been discussed in curved structures of 
superconductors~\cite{ya05} and superconducting weak links~\cite{ya05,rashedi}. 
To realize the spin current in experiments, we should consider simpler 
structures such as superconductor / insulator / superconductor (SIS) junctions of 
triplet superconductors.
Generally speaking, spin-triplet pairs are characterized by odd parity $p$- or $f$-wave
symmetry. In SIS junctions of such unconventional superconductors,
a midgap Andreev resonant state (MARS) governs the low energy transport~\cite{tanaka0}.
So far it is pointed out that the MARS drastically enhances the electric charge transport
but it suppresses the thermal transport~\cite{yokoyama}. 
Effects of the MARS on the spin transport is still an open question.
In contrast to spin-triplet pairs of $^3$He, Cooper pairs 
in superconductors have the charge degree of freedom which couples with
electromagnetic fields. This feature may also enables us to switch the spin current by
using electric fields and/or magnetic fields. 

In this paper, I calculate analytically the spin current in SIS junctions 
based on the mean-field theory of superconductivity. 
When the MARS is forming at the junction interface, the low-temperature anomaly 
appears in the spin current as well as it does in the charge current~\cite{barash,yt96-1}.
The spin current shows the Fraunhofer pattern in applied magnetic fields.
I also show that the applied bias-voltage across junctions causes 
the alternating current of spin. 

Throughout this paper, we take the unit of $\hbar=k_B=c=1$, where $k_B$ is the 
Boltzmann constant and $c$ denotes the speed of light. The vectors in the real and momentum space
are indicated by $\vec{\cdots}$ while those in the spin space are described by 
bold-italic characters.

\begin{figure}[thb]
\begin{center}
\includegraphics[width=7.5cm]{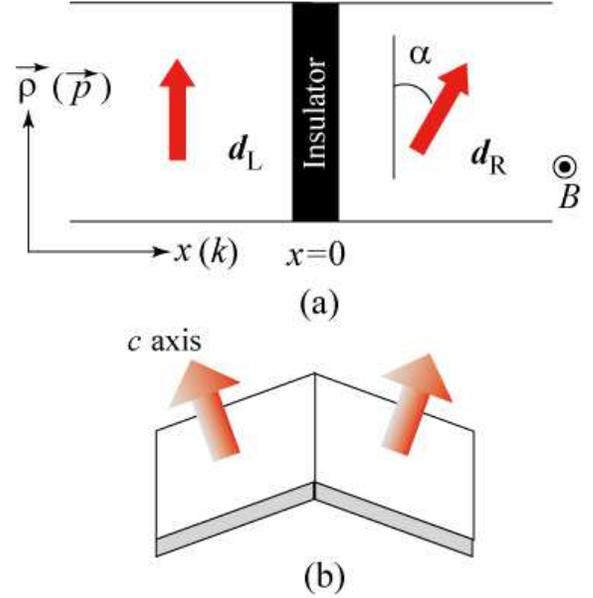}
\end{center}
\caption{ (Color online) (a) A schematic figure of a SIS junction.
The large arrows in superconductors represent $\boldsymbol{d}$ which is 
a unit vector in the spin space.
In (b), we propose a SIS junction using Sr$_2$RuO$_4$, where $\boldsymbol{d}$
is parallel to the $c$ axis of the crystal. 
}
\label{fig1}
\end{figure}

Electronic states in SIS junctions are described by the Bogoliubov-de Gennes (BdG) equation,
\begin{align}
&\int\!\! d\vec{r}\,' 
\!\!\left[\begin{array}{cc}
\hat{h}(\vec{r},\vec{r}\,') &
\hat{\Delta}(\vec{r},\vec{r}\,') \\
- \hat{\Delta}^\ast(\vec{r},\vec{r}\,') &
-\hat{h}^\ast(\vec{r},\vec{r}\,')
\end{array}\right]\!\!
\left[\begin{array}{c}
\hat{u}(\vec{r}\,')\\
\hat{v}(\vec{r}\,')
\end{array}\right] 
= E \!\!\left[\begin{array}{c}
\hat{u}(\vec{r}\, )\\
\hat{v}(\vec{r}\, )
\end{array}\right], \label{bdg}\\
&\hat{h}(\vec{r},\vec{r}\,')=\delta(\vec{r}-\vec{r}\,')
\left\{
-\frac{\vec{D}_{\vec{r}}^2}{2m} +V(\vec{r}\,)-\mu\right\}\hat{\sigma}_0,\label{h1}
\end{align}
where $\vec{D}_{\vec{r}\,}=\vec{\nabla}_{\vec{r}}-ie
\vec{A}_{\vec{r}}$ with $\vec{A}_{\vec{r}}$ being the vector potential, 
$\hat{\cdots}$ indicates $2\times 2$ matrix 
describing spin space, 
$\hat{\sigma}_0$ is the unit matrix
and $\mu$ is the Fermi energy. 
The insulating barrier at $x=0$ is described by $V_B\delta(x)$. 
In uniform superconductors, the pair potential is given by
\begin{align}
\hat{\Delta} \left( \vec{r},\vec{r}\,'\right)
=& \frac{1}{V_{vol}} \sum_{\vec{k}} \hat{\Delta}_{\vec{k}}
e^{i\vec{k} \cdot (\vec{r}-\vec{r}\,')}, \\
\hat{\Delta}_{\vec{k}}=&
 i \; \Delta_{\vec{k}}\;  \boldsymbol{d}_j \cdot \hat{\boldsymbol{\sigma}} \hat{\sigma}_2
e^{i\varphi_j}.
\end{align}
The vector $\vec{k}=(k,\vec{p}\,)$ represents 
a wave numbers on the Fermi surface (i.e., $k^2+\vec{p}^{\, 2}=k_F^2$), where
 $k$ and $\vec{p}$ are wave number in the direction of the current and that 
in the directions transverse to the current, respectively.
I consider a similar model as that in Ref.~\onlinecite{kastening}.
We assume that the two superconductors are identical to each
other except for the directions of the unite vectors $\boldsymbol{d}_j$ ($j=L$ or $R$) as 
shown in Fig.~\ref{fig1}(a), where
$\alpha$ is an orientation angle between $\boldsymbol{d}_L$ and 
$\boldsymbol{d}_R$.
In unconventional superconductor junctions, the characteristic behaviors of 
Josephson charge current are very sensitive to dependences of the pair potential 
on wavenumbers. When 
$\Delta_{k,\vec{p}}$ and $\Delta_{-k,\vec{p}}$ have opposite signs to each
other, the MARS forming at junction interfaces causes the anomalous charge transport 
at low temperatures~\cite{tanaka0,ya04}. In this paper, I consider the two typical 
situations: $\Delta_{-k,\vec{p}}=\nu \Delta_{k,\vec{p}}$ with $\nu=\pm 1$.
The MARS forms (does not form) for $\nu=-1$ ($\nu=1$).
Within $p$-wave symmetries, $\nu=1$ ($\nu=-1$) corresponds to $p_y$- ($p_x$-) wave 
symmetry.
At first I solve the BdG equation in the two superconductors independently and 
obtain the wave functions
\begin{align}
\boldsymbol{\Psi}_L(x,\vec{\rho}\,) =& \check{\chi}_L
\left[
\left(\begin{array}{c} \hat{u} \\ \hat{v} 
\end{array}\right)
e^{ikx} \hat{a}
+
\left(\begin{array}{c} \hat{v} \nu\\ 
\hat{u} \\  
\end{array}\right)
e^{-ikx} \hat{b} \right.\nonumber \\
&+
\left(\begin{array}{c} \hat{u} \\ 
\hat{v} \nu 
\end{array}\right)
 \left.e^{-ikx} \hat{A}
+
\left(\begin{array}{c} \hat{v} \\ 
\hat{u} \\  
\end{array}\right)
e^{ikx} \hat{B} 
\right] e^{i\vec{p}\cdot\vec{\rho}},\nonumber \\
\boldsymbol{\Psi}_R(x,\vec{\rho}\,) =& \check{\chi}_R
\left[
\left(\begin{array}{c} \hat{u} \\ \hat{v} 
\end{array}\right)
e^{ikx} \hat{C}
+
\left(\begin{array}{c} \hat{v} \nu \\ 
\hat{u} \\  
\end{array}\right)
e^{-ikx} \hat{D} \right]
\times e^{i\vec{p}\cdot\vec{\rho}},\nonumber\\
\check{\chi}_{j} =&\left(
\begin{array}{cc} e^{i\varphi_j/2}\hat{\sigma}_0 & \hat{0} \\
\hat{0} & e^{-i\varphi_j/2}\hat{\sigma}_0\end{array} \right),\nonumber
\end{align}
where $u (v)=\sqrt{(\omega_n+(-)\Omega)/(2\omega_n)}$, 
$\Omega=\sqrt{\omega_n^2+|\Delta_{\vec{k}}|^2}$, 
$\hat{u}=u\hat{\sigma}_0$, $\hat{v}= v {\hat{\Delta}_{\vec{k}}}/{|\Delta_{\vec{k}}|}$,
 and $\omega_n=(2n+1)T$ is the Matsubara frequency. 
The incident amplitudes of a quasiparticle in the electron (hole) branch
are denoted by diagonal $2\times 2$ matrix $\hat{a}$ ($\hat{b}$).
In $\boldsymbol{\Psi}_L$ ($\boldsymbol{\Psi}_R$),
$\hat{A}$ and $\hat{B}$ ($\hat{C}$ and $\hat{D}$) represent 
the amplitude of outgoing wave in the electron and hole branches, respectively.
Secondly I calculate the Andreev reflection
coefficients from the boundary conditions,
\begin{align}
\boldsymbol{\Psi}_L(0,\vec{\rho}\,)=&\boldsymbol{\Psi}_R(0,\vec{\rho}\,),\nonumber\\
\left.\frac{d}{dx}\boldsymbol{\Psi}_L(x,\vec{\rho}\,)\right|_{x\to 0}\!\!\!\!\!\!\!\!
=& \left.\frac{d}{dx}\boldsymbol{\Psi}_R(x,\vec{\rho}\,)\right|_{x\to 0}\!\!\!\!\!\!\!
- 2z_0k_F \boldsymbol{\Psi}_R(0,\vec{\rho}\,),\nonumber
\end{align}
with $z_0=V_B/(k_F/m)$.
The Andreev reflection coefficients 
are defied by the off-diagonal elements of the matrix relation below
\begin{equation}
\left(\begin{array}{c}
\hat{A} \\ \hat{B} \end{array}\right)
=\left(\begin{array}{cc}
\hat{r}_{ee} & \hat{r}_{eh}\\ 
\hat{r}_{he} & \hat{r}_{hh}\end{array}\right)
\left(\begin{array}{c}
\hat{a} \\ \hat{b} \end{array}\right).
\end{equation}
The calculated results of the Andreev reflection coefficients are summarized as
\begin{align}
\hat{r}_{he}=&\frac{uv\nu \hat{\Delta}^\dagger_{\vec{k}} \hat{K}}{\Xi  |\Delta_{\vec{k}}|}, \quad
\hat{r}_{eh}=\frac{uv \hat{K}^\ast \hat{\Delta}_{\vec{k}} }{\Xi |\Delta_{\vec{k}}|},\\
\hat{K}=&\left[(a_0c_0-a_1c_1\sin^2\!\!\alpha)\hat{\sigma}_0 +(a_0c_1-c_0a_1)
\boldsymbol{n}\!\cdot\!\hat{\boldsymbol{\sigma}}\right]/h_1h_2,\nonumber \\
a_0=&|r|^2 f_-^2(\nu g_+ \cos\alpha-f_+) +|t|^2(f_+ \nu \cos\alpha-g_+^\ast)h_1,\nonumber\\
a_1=&i\nu f_-\left[ |r|^2 g_-f_- + |t|^2h_1 \right],\nonumber\\
c_0 =& h_1\left\{\gamma - (1+\nu)\cos\alpha \right\}\nonumber\\ 
&-|r|^2 h_2\left\{ g_+ -u^2v^2\gamma - u^2v^2(1+\nu)\cos\alpha\right\},\nonumber\\
c_1=&-i(1-\nu) \left[h_1 + |r|^2h_2 u^2v^2\right],\nonumber\\
h_1=& u^4e^{i\varphi}+v^4e^{-i\varphi} - 2u^2v^2\cos\alpha, \nonumber\\ 
h_2=& 2( \cos\varphi - \nu \cos\alpha ),\nonumber\\
\Xi_\pm =& |r|^2 f_-^2 + |t|^2\left\{ 1-4u^2v^2\cos^2\left(\frac{\varphi\pm\alpha}{2}\right)
\right\}, \nonumber
\end{align}
with $f_\pm=u^2\pm \nu v^2$, $g_\pm = u^2e^{i\varphi} \pm v^2 e^{-i\varphi}\nu$,
$\gamma=e^{i\varphi}+e^{-i\varphi}\nu$, $\varphi=\varphi_L-\varphi_R$, 
$\boldsymbol{n}= \boldsymbol{d}_R \times \boldsymbol{d}_L$,
 and $\Xi=\Xi_+\Xi_-$. 
The normal transmission and reflection probabilities for a channel characterized by $\vec{p}$ 
are given by $|t|^2=k^2/(z_0^2k_F^2+k^2)$ and $|r|^2=z_0^2k_F^2/(z_0^2k_F^2+k^2)$, respectively.
Finally I obtain the spin current in SIS junctions based on the formula~\cite{ya05} 
\begin{align}
{\boldsymbol{J}}_s=- &\sum_{\vec{p}} \frac{T}{4} 
\sum_{\omega_n} \textrm{Tr} \left[
\frac{1}{\Omega}\left\{
\hat{\Delta}_{\vec{k}} \;\hat{r}_{he} \frac{\hat{\boldsymbol{\sigma}}}{2}
+  \hat{r}_{he} \;\hat{\Delta}_{\vec{k}} \frac{\hat{\boldsymbol{\sigma}}^\ast}{2}\right\}
 \right.\nonumber \\
&\left.
-  \frac{\nu}{\Omega}\left\{
\hat{r}_{eh}\; \hat{\Delta}_{\vec{k}}^\dagger \frac{\hat{\boldsymbol{\sigma}}}{2}
+ \hat{\Delta}_{\vec{k}}^\dagger \;\hat{r}_{eh} \frac{\hat{\boldsymbol{\sigma}}^\ast}{2}
\right\}\right].\label{sf2}
\end{align}
The electric Josephson current is also given by $\hat{\boldsymbol{\sigma}}/2 \to -e \hat{\sigma}_0$
in Eq.~(\ref{sf2})~\cite{ya01,furusaki}. 
The spin polarized in the direction of $\boldsymbol{n}\parallel \vec{z}$ flows through the 
SIS junction as shown in Fig.~\ref{fig1}(a).
The electric current $J_e$ and the $z$ component of the spin current ${J}_s^z$ 
in the $x$ direction result in
\begin{align}
\tilde{J_e}(\varphi,\alpha)=&\frac{J_e}{e\Delta_0} = \sum_{\vec{p}} \frac{|t|^2}{4} 
\frac{\Delta_{\vec{k}}}{\Delta_0} 
[ F_\nu(+) + F_\nu(-)], \label{je}\\
\tilde{J_s^z}(\varphi,\alpha)=&\frac{J_s^z}{\Delta_0/2} = \sum_{\vec{p}} \frac{|t|^2}{4} 
\frac{\Delta_{\vec{k}}}{\Delta_0} 
[ F_\nu(+) - F_\nu(-)],\label{js}\\
F_{\nu}(\pm) =& \frac{\sin(\varphi\pm\alpha)}{A_\nu(\pm)}\tanh
\left( \frac{\Delta_{\vec{k}} A_\nu(\pm)}{2T} \right), \\
A_\nu(\pm) = &\left\{
\begin{array}{cc}  \sqrt{1-|t|^2\sin^2\left(\frac{\varphi\pm\alpha}{2}\right)} & : \; 
\nu=1 \\
|t|\cos\left(\frac{\varphi\pm\alpha}{2}\right) & : \; \nu=-1
\end{array}\right.,
\end{align}
where $\Delta_0$ is the amplitude of the pair potential at zero temperature.
The summation in terms of $\vec{p}$ means the summation of the current over 
propagation channels.
\begin{figure}[thb]
\begin{center}
\includegraphics[width=7.5cm]{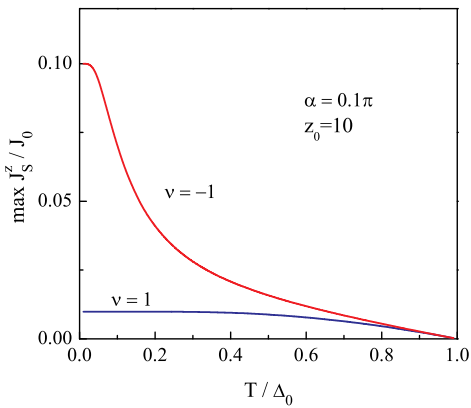}
\end{center}
\caption{ (Color online) The temperature dependence of the spin current for
$\alpha=0.1\pi$ and $z_0=10$. The results are normalized by the maximum 
value of the spin current at $z_0=0$ and $T=0$.
}
\label{fig2}
\end{figure}
It is easy to confirm that $J_s$ ($J_e$) is an odd function of $\alpha$ ($\varphi$). 
At $z_0=0$, $J_e$ and ${J}_s^z$ are independent of $\nu$ because
the transmission probability is unity (i.e., $|t|^2 = 1$). 
In what follows, we consider the low transmission limit (i.e., $|t|^2 \ll 1$).
We show the maximum amplitude of $\tilde{J}_s^z$ as a function of temperatures 
in Fig.~\ref{fig2} for $z_0=10$ and
$\alpha=0.1\pi$. The maximum values are calculated from the current-phase 
($\tilde{J}_s^z-\varphi$) relation and
are normalized by $J_0=\sum_{\vec{p}} |\Delta_{\vec{k}}|/\Delta_0 |\sin(\alpha/2)|$ 
which corresponds to the critical current at $z_0=0$ and $T=0$.
For $\nu=1$, the temperature dependence of the spin current is described by 
the Ambegaokar-Baratoff formula~\cite{ambegaokar} and the spin current
saturates at low temperatures as shown in Fig.~\ref{fig2}.
On the other hand for $\nu=-1$, the spin current increases rapidly with decreasing 
temperatures because the MARS forms at the interface.
The resonant transmission through the MARS causes the low-temperature
anomaly not only in the charge transport~\cite{yt96-1,barash} but also in the spin transport.
This is because the Andreev reflection carries both electric charges and spins as shown in 
Eq.~(\ref{sf2}). In fact, we find the relation 
$\tilde{J}_e(\alpha,\varphi)=\tilde{J}_s^z(\varphi,\alpha)$. 
Interchange of the angle in the gauge space $\varphi$ and the angle in the spin space $\alpha$ 
connects the two the supercurrents,
which implies the duality of the charge and spin.

In contrast to spin-triplet pairs of $^3$He, Cooper pairs in superconductors couple with 
the electromagnetic field through the charge degree of freedom of an electron.
As a result, the spin current is expected to be sensitive to electromagnetic fields.
Here I first consider the spin current in SIS junctions under magnetic fields.
It is well known that the electric Josephson critical current exhibits 
the Frounhofer pattern under magnetic fields. 
We apply the widely accepted way~\cite{barone} to Eq.~(\ref{js}) and calculate 
the critical spin current near the critical temperature as
\begin{align}
|J_s^z(\Phi)|=&|J_s^z(0)| 
\left| \sin\left( \frac{\pi\Phi}{\Phi_0} \right) \Big/ \left( \frac{\pi\Phi}{\Phi_0} \right),
\right|\label{fh}
\end{align}
where a magnetic field $B$ is parallel to the $z$ direction,
$\Phi=d_0WB$ with $d_0/2$ being a penetration depth of magnetic fields and 
$\Phi_0$ is the flux quantum.
The critical current of spin shows exactly the same Frounhofer pattern as 
that of the charge current.  For $\nu=-1$, the period of oscillations becomes 
$2\Phi_0$ in low temperatures due to the resonant transmission through the MARS. 
Next I discuss the spin current under applied bias-voltage. 
The ac Josephson effect is also well known in SIS junctions of spin-singlet
superconductors. The macroscopic phase obeys the Josephson's equation of motion
$ {\partial \varphi}/{\partial t} = 2eV_{\textrm{bias}}$
under the bias-voltage $V_{\textrm{bias}}$ across the junction.
The spin current under the bias-voltage near the critical temperature becomes
\begin{equation}
\tilde{J}_s^z \propto \cos(  2eV_{\textrm{bias}}t + \varphi_0)\sin\alpha. \label{ac}
\end{equation}
The bias-voltages generate the alternating spin current in SIS junctions.
The gauge coupling of the Cooper pair results in Eqs.~(\ref{fh}) and (\ref{ac}) 
which represent the specific properties of the spin current in superconductor junctions.
In addition, it may be possible to control the spin current by using these properties. 

To observe the spin current in superconducting materials or junctions, we have to 
overcome two difficulties in experiments: the generation and detection of the spin current.
To generate the spin current, we propose a SIS junction of the spin-triplet 
superconductor~\cite{maeno} Sr$_2$RuO$_4$ as shown in Fig.~\ref{fig1}(b). 
The bending junction is required to have the spatial gradient of $\boldsymbol{d}$. 
In Sr$_2$RuO$_4$, $\boldsymbol{d}$ 
homogeneously aligns in the parallel direction to the $c$ axis of the crystal lattice. 
The fabrication of such bending SIS junction may be possible because the crystal growth 
rate within the $ab$ plane is much faster than that in the $c$ axis. The grain boundary 
separates the two thin films of single crystal. 
At present, the chiral $p_x\pm ip_y$-wave symmetry
is the promising candidate of the pairing symmetry in Sr$_2$RuO$_4$.
The formation of the MARS depends on incident angles of a quasiparticle into junction 
interface. 
As a result, the MARS is weakly forming at the interface.
The spin current has an intermediate quality between the spin current with $\nu=1$ and 
that with $\nu=-1$. 
Although Eqs.~(\ref{js}), (\ref{fh}) and (\ref{ac}) are derived in the ballistic junctions,
these behaviors persist even in realistic disordered junctions.
This is because the MARS causes the anomalous Josephson effect~\cite{ya06-1} in the $p$-wave
symmetry. The electromagnetic fields modulate the spin current as
discussed in Eqs.(\ref{fh}) and (\ref{ac}). 
At present, unfortunately, there is no good idea of directly observing the spin flow.
Actually, the spin Hall effect~\cite{murakami} is confirmed by the spin accumulation 
which is a result of the spin flow. The application of the spin-orbit device~\cite{nikolic} 
may be possible to detect the spin current in superconducting junctions.

In summary, I have studied the spin supercurrent in superconductor / insulator / superconductor
junctions, where superconductors are in spin-triplet unitary states.
 On the basis of the current formula, I calculate spin and charge current analytically 
from the Andreev reflection coefficients of the junctions.
The resonant transmission through the midgap Andreev resonant state 
causes the low-temperature anomaly of the spin current. 
The gauge coupling of the Cooper pairs with electromagnetic field
results in the Frounhofer pattern of the spin current under magnetic fields 
and the alternating spin flow in the presence of bias-voltages.
The application of the obtained results to realistic junctions consisting
of Sr$_2$RuO$_4$ is briefly discussed.

The author acknowledges helpful discussions with Y.~Maeno, Y.~Tanaka, S.~Kashiwaya, 
A.~A.~Golubov, J.~Aarts, K.~Nagai and S.~Kawabata.
 This work has been partially supported by Grant-in-Aid for the 21st Century
COE program on "Topological Science and Technology",
and Grant-in-Aid for Scientific Research on Priority Area
"Physics of new quantum phases in superclean materials"
(Grant No. 18043001) from The Ministry of Education, Culture, Sports, 
Science and Technology of Japan.


\begin{thebibliography}{} 


\bibitem{leggett}
A.~J.~Leggett, Rev. Mod. Phys. \textbf{47}, 331 (1975).

\bibitem{bunkov}
Yu.~M.~Bunkov, Chapter~2 in \textit{Progress in Low Temperature Physics} XIV
Edited by W.~P.~Halperin, North-Holland (1995).

\bibitem{vuorio} M.~Vuorio, J. Phys. C \textbf{9}, L267 (1976).

\bibitem{fomin} I.~A.~Fomin, JETP Lett. \textbf{40}, 1037 (1984).

\bibitem{ya05} Y.~Asano, Phys. Rev. B \textbf{72}, 092508 (2005).

\bibitem{rashedi} G.~Rashedi and Yu.~A.~Kolesnichenko, cond-mat/0501211.

\bibitem{tanaka0} Y.~Tanaka and S.~Kashiwaya, Phys. Rev. Lett. \textbf{74}, 3451 (1995);
S.~Kashiwaya and Y.~Tanaka, Rep. Prog. Phys. \textbf{63}, 1641 (2000).

\bibitem{yokoyama} T.~Yokoyama, Y.~Tanaka, A.~A.~Golubov, and Y.~Asano,
Phy. Rev. B \textbf{72}, 214513 (2005).

\bibitem{yt96-1} Y.~Tanaka and S.~Kashiwaya, Phys. Rev. B \textbf{53}, R11957 (1996).

\bibitem{barash} Y.~S.~Barash, H.~Burkhardt, and D.~Rainer, 
Phys. Rev. Lett. \textbf{77}, 4070 (1996).

\bibitem{kastening} B.~Kastening, D.~K.~Morr, D.~Manske, and K.~Bennemann,
Phys. Rev. Lett. \textbf{96}, 047009 (2006).

\bibitem{ya04} Y.~Asano, Y.~Tanaka, and S.~Kashiwaya, Phys. Rev. B, \textbf{69}, 134501 (2004).

\bibitem{ya01} Y.~Asano, Phys. Rev. B \textbf{64}, 224515 (2001).
\bibitem{furusaki} A.~Furusaki and M.~Tsukada, Solid State Commun. \textbf{78}, 299 (1991).

\bibitem{ambegaokar} V.~Ambegaokar and A.~Baratoff, Phys. Rev. Lett. \textbf{10}, 486 (1963). 

\bibitem{barone} A.~Barone and G.~Paterno, "\textit{Physics and Applications of the Josephson 
Effect}", John Wiley \& Sons, New York (1982).



\bibitem{maeno} Y.~Maeno, H.~Hashimoto, K.~Yoshida, S.~Nishizaki, T.~Fujita,
 J.~G.~Bednorz, and F.~Lichtenberg, Nature \textbf{372}, 532 (1994).


 
\bibitem{ya06-1}Y.~Asano, Y.~Tanaka, and S.~Kashiwaya, Phys. Rev. Lett. \textbf{96}, 097007 (2006).

\bibitem{murakami} S.~Murakami, N.~Nagaosa, and S.~C.~Zhang, Science \textbf{301}, 1348 (2003).

\bibitem{nikolic} B.~K.~Nikolic, S.~Souma, L.~P.~Zarbo, and J.~Sinova,
Phys. Rev. Lett. \textbf{95}, 046601 (2005).


\end{thebibliography}
\end{document}